\documentclass[a4paper, amsfonts, amssymb, amsmath, reprint, showkeys, nofootinbib, twoside, onecolumn, notitlepage]{revtex4-1}

\usepackage[centering,hmargin=2cm,vmargin=2cm,lmargin=1.7cm,rmargin=1.7cm]{geometry}

\def\apj{{ApJ}}                 

\def\mnras{{MNRAS}}             
\def\prc{{Phys.~Rev.~C}}        
\def\prd{{Phys.~Rev.~D}}        
\def\prl{{Phys.~Rev.~Lett.}}    
\usepackage{amsmath,amstext}
\usepackage[T1]{fontenc}
\usepackage{amssymb}
\usepackage{graphicx}
\usepackage{ae,aecompl}

\DeclareFontFamily{OT1}{pzc}{}
\DeclareFontShape{OT1}{pzc}{m}{it}{<-> s * [1.10] pzcmi7t}{}
\DeclareMathAlphabet{\mathpzc}{OT1}{pzc}{m}{it}

\usepackage{hyperref}
\usepackage{amsmath}
\usepackage{amssymb}
\usepackage{mathtools}
\usepackage{bm}
\usepackage{cleveref}
\usepackage{tensor}
\usepackage{braket}
\usepackage{enumitem}
\usepackage{mhchem}
\usepackage{amsthm}
\usepackage{nccmath}
\usepackage{color}
\usepackage{mathrsfs}
\usepackage{pgfplots}
\pgfplotsset{compat=1.18}

\newcommand{\spc}{\quad \quad \quad}

\newcommand{\F}{{\mathfrak{F}}}

\def\be{\begin{equation}}
\def\ee{\end{equation}}
\def\beq{\begin{eqnarray}}
\def\eeq{\end{eqnarray}}

\begin{document}
\title{Relativistic heat conduction in the large-flux regime}
\author{L.~Gavassino}

\affiliation{
Department of Mathematics, Vanderbilt University, Nashville, TN, USA
}

\begin{abstract}
We propose a general procedure for evaluating, directly from microphysics, the constitutive relations of heat-conducting fluids in regimes of large fluxes of heat. Our choice of hydrodynamic formalism is Carter's two-fluid theory, which happens to coincide with \"{O}ttinger's GENERIC theory for relativistic heat conduction. This is a natural framework, as it should correctly describe the relativistic ``inertia of heat'' as well as the subtle interplay between reversible and irreversible couplings. We provide two concrete applications of our procedure, where the constitutive relations are evaluated respectively from maximum entropy hydrodynamics and Chapman-Enskog theory.
\end{abstract} 

\maketitle

\section{Introduction}\label{intro}

The most widespread theory for relativistic dissipation is the Israel-Stewart theory \cite{Israel_Stewart_1979,Hishcock1983}, which was proven to be very effective in modeling viscosity and heat conduction in relativistic gases \cite{Denicol2012Boltzmann} and liquids \cite{GavassinoGENERIC2023}. The rationale of the Israel-Stewart theory is rooted in Extended Irreversible Thermodynamics \cite{Jou_Extended}, which posits that the dissipative fluxes, like the heat flux $q$, should be treated as non-equilibrium thermodynamic variables. This allows one to define a non-equilibrium entropy density $s(q^2)$, which can be expanded to second order in the flux:
\begin{equation}\label{suissius}
    s(q^2)=s_{\text{eq}}-\dfrac{\beta_1}{2T} q^2 +\mathcal{O}(q^4) \, .
\end{equation}
Using the second law of thermodynamics as a guiding principle, one can then derive some dissipative equations of motion for the fluxes, which resemble Catteno's model: $\tau \dot{q}+q \propto \text{``local gradients''}$ \cite{cattaneo1958}. This model has indeed been shown to be consistent with the kinetic theory of gases \cite{Denicol_Relaxation_2011,WagnerGavassino2023} and the rheological theory of liquids \cite{Frenkel_book,BAGGIOLI20201,GavassinoCarterGeneric2023}. 

It is natural to ask whether we can extend the Israel-Stewart framework beyond the second order in dissipative fluxes\footnote{In this article, we interpret the Israel-Stewart framework as a formulation of transient hydrodynamics, and not as a gradient expansion, see \cite{WagnerGavassino2023}, interpretation (iii). This means that $q$ is a dynamical effective field, which parameterizes the displacement of the fluid from local equilibrium. In the case of heat conduction, Israel-Stewart hydrodynamics reduces to the M1 closure scheme \cite{Levermore1981,Sadowski2013,GavassinoRadiazione}.}. This question was addressed in the literature for certain dissipative processes, and the resulting formalism seems to depend on the flux under consideration. For the bulk stress, the extension of Israel-Stewart beyond quadratic order can be identified with Hydro+ \cite{Stephanov:2017ghc,BulkGavassino,GavassinoFarbulk2023}. For the shear stress, the extension is called ``anisotropic hydrodynamics'' \cite{Strickland:2014pga,Alqahtani:2017mhy}. In this work, we focus on heat conduction, which is probably the least understood case. 

The two most promising extensions of the Israel-Stewart theory for heat conduction beyond quadratic order are Carter's multifluid theory \cite{noto_rel,carter1991,lopez2011} (which treats heat as a carrier of inertia) and the GENERIC theory \cite{OttingerSoloHeat1998}. These were recently proven to be the same mathematical system of equations \cite{GavassinoCarterGeneric2023} (just written in different variables). Both approaches define a heat-flux-dependent equation of state, similarly to \eqref{suissius}, which can be in principle extrapolated to large values of heat flux. However, to date, no practical procedure has been proposed to compute such an equation of state from microscopic models. This article aims to propose such a procedure.

Throughout the article, we adopt the metric signature $(-,+,+,+)$, and work in natural units: $c=k_B=\hbar=1$.

\section{Mathematical structure of the GENERIC-Multifluid theory for heat conduction}

First, let us analyze the theory of \citet{OttingerSoloHeat1998} for relativistic heat conduction. Such theory arises from direct application of the GENERIC framework \cite{Grmela1997} to relativistic conductive fluids. Since this same theory can also be derived within Carter's multifluid framework \cite{GavassinoCarterGeneric2023}, we will refer to it as the GENERIC-Multifluid (GM) theory.

\subsection{Non-equilibrium thermodynamics}\label{noneqther}

The fields of the GM theory are, by assumption, $\Psi=\{n,u^\mu,w_\mu\}$. The first two may be interpreted as the rest-frame baryon density and the (Eckart frame \cite{Kovtun2019}) flow velocity. The covector $w_\mu$ is an effective non-equilibrium field, usually called ``thermal momentum'' \cite{lopez2011}. The non-equilibrium temperature is \textit{defined} to be $T=-u^\mu w_\mu$. If the fluid is in local thermodynamic equilibrium, $w^\mu$ must be parallel to $u^\mu$ (by isotropy). It follows that the non-negative definite scalar $w^\mu w_\mu{+}T^2$ can be interpreted as a measure of how far from local equilibrium the fluid is. This motivates (in agreement with Extended Irreversible Thermodynamics \cite{Jou_Extended}) the introduction of a non-equilibrium free-energy density $\mathfrak{F}(T,n,w^\mu w_\mu)$, which has an absolute minimum at $w^\mu w_\mu =-T^2$, for fixed values of $n$ and $T$ \cite{Termo,Callen_book}. We define the non-equilibrium entropy density $s$ and chemical potential $\mu$ from the following differential:
\begin{equation}\label{free}
    d\F = -s \, dT+ \mu \, dn +\dfrac{\sigma}{2} d(w^\mu w_\mu {+}T^2) \, .
\end{equation}
At equilibrium, the thermodynamic coefficient $\sigma$ is necessarily positive (for $\F$ to be in a minimum \cite{Callen_book}). Additionally, we can define the thermodynamic energy density $\varepsilon=\F+Ts$ and the thermodynamic pressure $P=-\F+\mu n$, as in standard thermodynamics \cite{landau5}. The thermodynamic identities below follow directly from the above definitions:
\begin{equation}\label{thermo}
    \begin{split}
d\varepsilon ={}& T \, ds +\mu \, dn +\dfrac{\sigma}{2} d(w^\mu w_\mu {+}T^2) \, , \\
dP ={}& s \, dT + n \, d\mu - \dfrac{\sigma}{2} d(w^\mu w_\mu {+}T^2) \, , \\
\varepsilon{+}P={}& Ts+\mu n \, . \\
    \end{split}
\end{equation}

\subsection{Hydrodynamic constitutive relations}

The effective fields $\Psi$ are not observable. They are just mathematical degrees of freedom that we use to parameterize the macroscopic state of the system. Within relativistic hydrodynamics, the relevant physical observables are the following fluxes: $T^{\mu \nu}$ (the stress-energy tensor), $s^\mu$ (the entropy current), and $n^\mu$ (the baryon current). Thus, we need some formulas to express these fluxes in terms of the effective fields $\Psi$. Such formulas are usually referred to as constitutive relations. For the GM theory, the constitutive relations are postulated to be
\begin{equation}\label{constitutiverelations}
    \begin{split}
T^{\mu \nu}={}& P g^{\mu \nu}+(\varepsilon+P-\sigma T^2)u^\mu u^\nu +\sigma w^\mu w^\nu \, , \\
s^\mu ={}& (s-\sigma T)u^\mu +\sigma w^\mu \, , \\
n^\mu ={}& n u^\mu \, . \\
    \end{split}
\end{equation}
These are just the most natural constitutive relations that one can write working in the Eckart frame, i.e. assuming that $n^\mu \propto u^\mu$, $n=-n^\mu u_\mu$, and $\varepsilon=T^{\mu \nu}u_\mu u_\nu$. Indeed, defined the heat flux vector
\begin{equation}\label{heatflux}
    q^\mu =\sigma T (w^\mu {-}Tu^\mu) \, ,
\end{equation}
which satisfies the orthogonality condition $u_\mu q^\mu =0$, we can rewrite the constitutive relations \eqref{constitutiverelations} as follows:
\begin{equation}\label{constitutiverelations2}
    \begin{split}
T^{\mu \nu}={}& P g^{\mu \nu}+(\varepsilon+P)u^\mu u^\nu+u^\mu q^\nu +q^\mu u^\nu  +\dfrac{q^\mu q^\nu}{T^2 \sigma} \, , \\
s^\mu ={}& s u^\mu +\dfrac{q^\mu}{T} \, , \\
n^\mu ={}& n u^\mu \, . \\
    \end{split}
\end{equation}
These can be interpreted as non-perturbative generalizations of the Israel-Stewart constitutive relations \cite{Hishcock1983,OlsonRegular1990,PriouCOMPAR1991} (in the Eckart frame). It should be kept in mind that all the thermodynamic variables may depend on the heat flux in a fully non-linear manner. Thus, the present theory is in principle applicable in regimes with large fluxes of heat.

\subsection{Consistency with relativistic thermodynamics}

Let us verify that the above theory is consistent with the principles of relativistic thermodynamics, in the Van-Kampen-Israel formulation \cite{VanKampen1968,Israel_Stewart_1979,Israel_2009_inbook,GavassinoTermometri,GavassinoLorentzInvariance2022}. Using equations \eqref{thermo} and \eqref{constitutiverelations}, one can easily prove the following identities:
\begin{equation}\label{gibbs}
    \begin{split}
Ts^\mu ={}& Pu^\mu -\mu n^\mu -u_\nu T^{\nu \mu} \, ,\\
Tds^\mu ={}& -\mu dn^\mu -u_\nu dT^{\nu \mu}-2\sigma u^{[\mu}w^{\nu]}dw_\nu \, . \\
    \end{split}
\end{equation}
The first equation is Israel's covariant Euler relation \cite{Israel_2009_inbook}. Note that, while in general this is an equilibrium identity, in the GM theory it happens to hold also in the presence of a heat flux. The second equation coincides with Israel's covariant Gibbs relation if and only if $w^\mu=Tu^\mu$. This implies that the fluid is in local thermodynamic equilibrium if and only if $q^\mu=0$ [see equation \eqref{heatflux}], i.e., there is no flow of heat across the fluid. Thus, the theory is indeed consistent with the principles of Van-Kampen-Israel thermodynamics.

It is also straightforward to verify that the GM theory describes a multifluid of Carter \cite{Carter_starting_point,Termo}. In fact, if we express the free energy as a function $\F(w_\alpha,n^\alpha,g^{\alpha \beta})$, we have the following partial derivatives:
\begin{equation}
\dfrac{\partial \F}{\partial w_\alpha} \bigg|_{n^\beta,g^{\beta \gamma}} = s^\alpha \, , \spc
\dfrac{\partial \F}{\partial n^\alpha} \bigg|_{w_\beta,g^{\beta \gamma}} = -\mu u_\alpha +\dfrac{s{-}\sigma T}{n}(w_\alpha{-}Tu_\alpha) \, , \spc 
2\dfrac{\partial \F}{\partial g^{\alpha\beta}} \bigg|_{n^\beta, w_\gamma} = T_{\alpha \beta}-Pg_{\alpha \beta} \, ,
\end{equation}
which are consistent with Carter's theory in the generating function formulation \cite{GavassinoKhalatnikov2022}. This implies that, as long as the second law of thermodynamics is respected, and the equation of state for $\F$ is prescribed in accordance with the requirements listed in \cite{GavassinoStabilityCarter2022}, the GM theory is linearly causal \cite{GavassinoCausality2021} and covariantly stable \cite{GavassinoGibbs2021}, both dynamically and thermodynamically \cite{PrigoginebookModernThermodynamics2014,Pathria2011}.

\subsection{Equations of motion}

To complete the theory, we need to prescribe some equations of motion for the fields $\Psi=\{n,u^\mu,w_\mu \}$. Since the algebraic degrees of freedom are $8$, we need $8$ independent equations of motion. Out of these, $5$ are the conservation laws $\nabla_\mu n^\mu=0$ and $\nabla_\mu T^{\mu \nu}=0$. The remaining $3$ are derived to guarantee consistency with the principles of GENERIC \cite{Grmela1997}. The simplest equation of motion fulfilling all the requirements is \cite{OttingerSoloHeat1998}
\begin{equation}\label{eom}
    u^\mu (\nabla_\mu w_\nu -\nabla_\nu w_\mu)=-\dfrac{1}{\tau} (w_\nu-Tu_\nu) \, ,
\end{equation}
where $\tau(T,n,w^\mu w_\mu)>0$ can be interpreted as the relaxation time. In \eqref{eom}, there are only $3$ independent equations, since contraction of both sides with $u^\nu$ returns a trivial identity ``$0{=}0$''. Consistency with GENERIC automatically entails consistency with the Onsager-Casimir principle \cite{Grmela2014,GavassinoCasimir2022}, and with the second law of thermodynamics. Indeed, with the aid of the second equation of \eqref{gibbs}, we can explicitly evaluate the entropy production rate:
\begin{equation}\label{secondlaw}
    T \nabla_\mu s^\mu = \dfrac{\sigma}{\tau} (w^\mu w_\mu +T^2) \, ,
\end{equation}
which is non-negative definite for arbitrary values of $\Psi$.

To get a better insight into the physical content of equations \eqref{eom} and \eqref{secondlaw}, we can express them in terms of the heat flux vector \eqref{heatflux}. The result is
\begin{equation}\label{grippont}
    \begin{split}
    \tau \sigma T \mathcal{L}_u\bigg( \dfrac{q}{\sigma T}\bigg)_\nu \! \! \! +q_\nu ={}& -\tau \sigma T (g\indices{^\mu _\nu}{+}u^\mu u_\nu)(T u^\lambda \nabla_\lambda u_\mu {+}\nabla_\mu T) \, , \\
T\nabla_\mu s^\mu ={}& \dfrac{q^\mu q_\mu}{\tau \sigma T^2} \, .\\
    \end{split}
\end{equation}
The consistency with Israel-Stewart theory \cite{Israel_Stewart_1979,Hishcock1983} in the limit of small heat fluxes is evident. The Lie derivative $\mathcal{L}_u$ in the first equation automatically accounts for the coupling with the vorticity predicted by kinetic theory \cite{Denicol2012Boltzmann}, and it guarantees that $q^\nu$ remains orthogonal to $u^\nu$ at all times.

\section{Evaluation of the constitutive relations from microphysics}

In the previous section, we outlined a general hydrodynamic framework for describing relativistic heat conduction non-perturbatively. Now we need a procedure for computing the non-equilibrium equation of state $\F(n,u^\mu,w_\mu)$ from microphysics. The main difficulty is that the field $w_\mu$ doesn't have a straightforward physical interpretation. Indeed, even $T$ itself is not clearly defined (out of equilibrium \cite{Kovtun2019}). This may open the doors to all sorts of ambiguities when trying to connect hydrodynamics with other levels of description, like kinetic theory. Here, we present a simple (and rigorous) procedure that allows one to circumvent all interpretative difficulties, and to evaluate $\F$ unambiguously.

\subsection{General strategy}\label{genna}

Pick a spacetime event $\mathscr{P}$, and move to the local rest frame of the fluid. Align the $x^1$ axis with the heat flux vector $q^\mu(\mathscr{P})$. Then, the constitutive relations \eqref{constitutiverelations} and \eqref{constitutiverelations2} can be expressed in components as follows:
\begin{equation}\label{struzzi}
    T^{\mu \nu}(\mathscr{P})=
    \begin{bmatrix}
    \varepsilon & \sigma T w & 0 & 0 \\
    \sigma T w & P{+}\sigma w^2 & 0 & 0 \\
    0 & 0 & P & 0 \\
    0 & 0 & 0 & P \\
    \end{bmatrix}=
    \begin{bmatrix}
    \varepsilon & q & 0 & 0 \\
    q & P_L & 0 & 0 \\
    0 & 0 & P_T & 0 \\
    0 & 0 & 0 & P_T \\
    \end{bmatrix} \, ,
\end{equation}
where $w=\sqrt{w^\mu w_\mu {+}T^2}$ is the non-equilibrium excursion, $q=\sqrt{q^\mu q_\mu}$ is the heat flux magnitude, $P_L=P+q^2/(T^2 \sigma)$ is the longitudinal pressure, and $P_T=P$ is the transversal pressure. Comparing the two matrices above, and recalling the second law \eqref{secondlaw}, we obtain
\begin{equation}\label{silo}
\dfrac{w}{T} = \dfrac{P_L{-}P_T}{q} \, , \spc
T^2 \sigma = \dfrac{q^2}{P_L{-}P_T} \, , \spc
 T =\dfrac{P_L{-}P_T}{\tau \nabla_\mu s^\mu} \, . 
\end{equation}
If one has a microscopic model for the heat flux (e.g. from kinetic theory), they can evaluate $q$, $P_L$, $P_T$, $\tau$, and $\nabla_\mu s^\mu$ explicitly. Note that there is no ambiguity in the kinetic definition of each of these quantities. Thus, there is no ambiguity over the exact values of $w$, $\sigma$, and $T$ for a given kinetic state. Varying the state, we can reconstruct the function $\sigma=\sigma(T,n,w^2)$. Additionally, the equilibrium free energy density $\F_{\text{eq}}(T,n)=\F(T,n,w^2{=}0)$ is known from statistical mechanics. Therefore, from \eqref{free}, one can finally compute the non-equilibrium free energy
\begin{equation}\label{farfree}
    \F(T,n,w^\mu w_\mu) = \F_{\text{eq}}(T,n) + \dfrac{1}{2} \int_0^{w^\mu w_\mu+T^2} \! \! \! \sigma(T,n,w^2) d(w^2) \, .
\end{equation}
Once $\F$ is known, all the constitutive relations can be computed through partial differentiation.

\subsection{Two simple examples}\label{ortega}

Suppose that heat is transported by a single branch of quasi-particle excitations, which have a long mean free path and carry zero net baryon number (so their motion does not modify the value of $u^\mu$). For simplicity, we assume that the variable $T$, defined in \eqref{silo}, fully characterizes the energy distribution of such excitations. It follows that the non-equilibrium excursion $w$ only affects the angular distribution of the excitation momenta, but not the magnitude of the momenta. Thus, we can express the heat flux and the pressure anisotropy in the following form:
\begin{equation}\label{finalmente}
    \begin{split}
q={}& v_{\text{ch}}(T,n) \, \mathfrak{R}(T,n) \int_{-1}^{+1} f(\cos\theta) \, \cos \theta \,  d (\cos\theta) \, , \\
P_L{-}P_T={}&  \mathfrak{R}(T,n) \int_{-1}^{+1} f(\cos\theta) \, \dfrac{3(\cos \theta)^2{-}1}{2} \, d (\cos\theta) \, , \\
    \end{split}
\end{equation}
where $\mathfrak{R}(T,n)$ is the average stress content of the excitation branch, $v_{\text{ch}}(T,n)$ is the characteristic speed of the branch, and $f(\cos \theta)$ is the (normalized) angular distribution of the momenta in the branch\footnote{Note that the fluid possesses other excitation branches, with shorter mean free path (some of which carry net baryon number). All these other excitations manage to thermalize (being short-lived). Hence, $\mathfrak{R}$ does not describe the totality of the stress trace $2P_T+P_L$. Instead, the stress trace can be decomposed as $\mathfrak{R}+3\mathfrak{C}$, where $\mathfrak{C}$ is an isotropic piece. At this stage, we don't need to model $\mathfrak{C}$ explicitly, because it does not contribute to the difference $P_L-P_T$, being isotropic.}. To determine $f$, we need a kinetic model for flux-limited diffusion. There are two popular proposals in the literature \cite{LEVERMORE1984}. The first, due to Minerbo \cite{MINERBO1978}, postulates that the angular distribution $f$ should maximize the entropy for the given value of heat flux $q$. The second, due to Levermore \cite{Levermore1979,Levermore1981}, is an approximate solution of the Boltzmann equation for the long-lived excitations, with a derivation that goes back to Chapman and Enskog \cite{ChapmanCowling}. Both approaches lead to a formula for $f$ that depends on a free parameter $Z \in [0,+\infty]$. The exact expressions for $f$ are provided below:
\begin{flalign}
\text{Minerbo:} && \label{twogang} f(x) &= \dfrac{Ze^{Zx}}{2\sinh Z} \;,&\\
\text{Levermore:} && \label{threegang} f(x) &= \dfrac{1}{2Z(\coth Z {-}x)} \;.
\end{flalign}
 When $Z{=}0$, the distribution is isotropic. When $Z{=}+\infty$, all the excitations travel in the direction of the heat flux. In appendix \ref{AAA}, we sketch the derivation of \eqref{finalmente}-\eqref{threegang}. Both models lead to the same prescription for the heat flux as a function of $Z$, namely
$q{=}v_{\text{ch}}\mathfrak{R} \, (\coth Z{-}1/Z)$, but predict different pressure anisotropies (see Fig. \ref{fig:qP}, left panel):
\begin{figure}
\begin{center}
\includegraphics[width=0.42\textwidth]{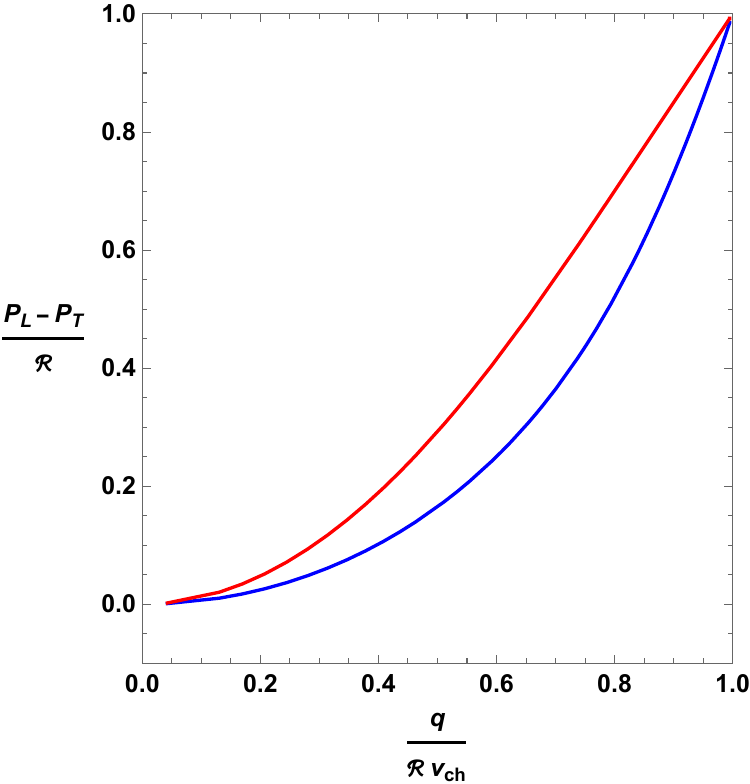}
\includegraphics[width=0.42\textwidth]{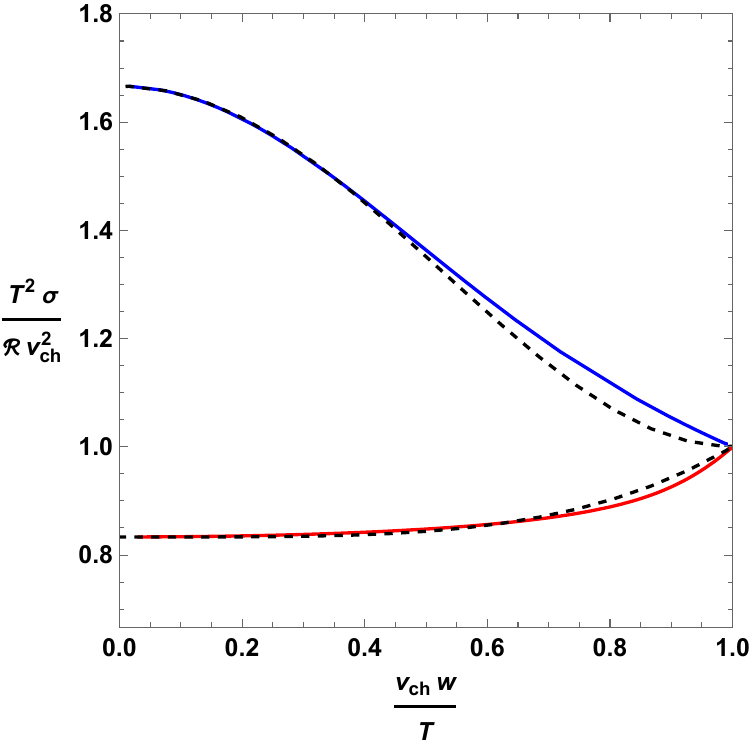}
	\caption{Non-equilibrium thermodynamics of heat conducting fluids according to the Minerbo prescription (blue) and the Levermore prescription (red). Left panel: Pressure anisotropy as a function of the heat flux. Right panel: Transport coefficient $\sigma$ as a function of the non-equilibrium excursion $w$. The dashed lines represent the polynomial fits.}
	\label{fig:qP}
	\end{center}
\end{figure}
\begin{flalign}
\text{Minerbo:} && \label{twogang2} \dfrac{P_L{-}P_T}{\mathfrak{R}} &= \dfrac{Z^2{-}3Z\coth Z{+}3}{Z^2} \;,&\\
\text{Levermore:} && \label{threegang2} \dfrac{P_L{-}P_T}{\mathfrak{R}} &= \dfrac{3\coth Z (Z\coth Z{-}1)}{2Z}-\dfrac{1}{2} \;.
\end{flalign}
Plugging these formulas into \eqref{silo}, we obtain two alternative constitutive relations $\sigma(w)$, which are plotted in figure \ref{fig:qP}, right panel. There is no analytical expression because, in both cases, the relation $\sigma(w)$ is given in a parametric form, $\{w(Z),\sigma(Z)\}$, and the dependence of $w$ on $Z$ does not admit an analytic inverse. However, we can fit the relations using a polynomial approximation. Below, we report a good compromise between analytical simplicity and accuracy (see dashed lines in figure \ref{fig:qP}):
\begin{flalign}
\text{Minerbo:} && \label{twogangcabba} \dfrac{T^2 \sigma}{\mathfrak{R} v_{\text{ch}}^2} &= \dfrac{5}{3}-\dfrac{3}{2} \bigg( \dfrac{v_\text{ch} w}{T}\bigg)^2 + \bigg( \dfrac{v_\text{ch} w}{T}\bigg)^4-\dfrac{1}{6} \bigg( \dfrac{v_\text{ch} w}{T}\bigg)^6 \;,&\\
\text{Levermore:} && \label{threegangcabba} \dfrac{T^2 \sigma}{\mathfrak{R} v_{\text{ch}}^2} &= \dfrac{5}{6}+\dfrac{1}{6}\bigg( \dfrac{v_\text{ch} w}{T}\bigg)^4 \;.
\end{flalign}
These approximations are designed to be very accurate up to $v_{\text{ch}}w/T \sim 0.5$. At larger heat fluxes, the accuracy is slightly lower ($\sim 5$\% error). However, at maximum heat flux, namely for $v_{\text{ch}}w/T = 1$, the polynomial approximation becomes exact. As a consistency check, we note that, in the limit of small $q$, the conductivity coefficient $\kappa=\tau \sigma T$, as predicted by equation \eqref{grippont}, has the correct scaling \cite{Tritt} for both models:
\begin{equation}
    \kappa \propto \dfrac{\tau \, \mathfrak{R} \, v_{\text{ch}}^2}{T} \, .
\end{equation}
Using equation \eqref{farfree}, we finally obtain the non-equilibrium free energy density for both models (see figure \ref{fig:freeee}):
\begin{figure}
\begin{center}
\includegraphics[width=0.50\textwidth]{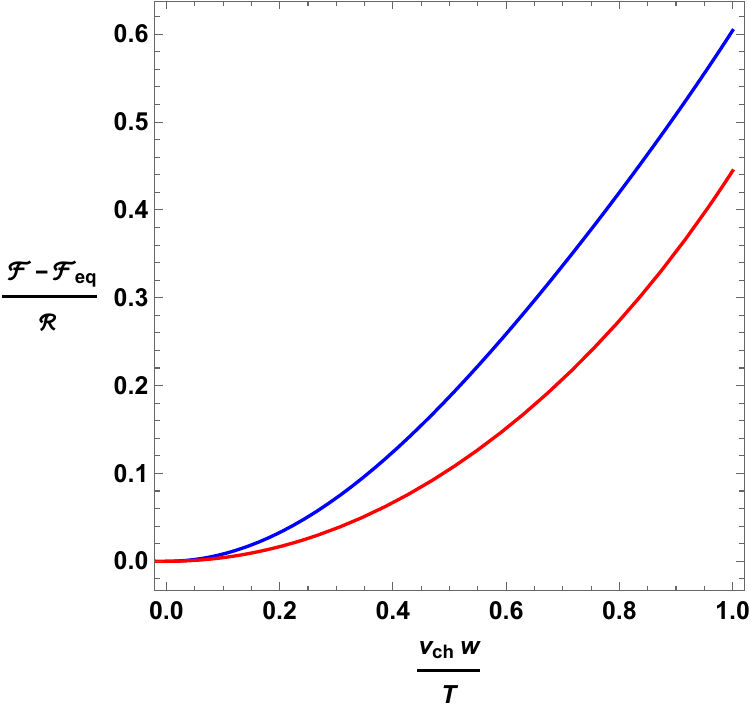}
	\caption{Non-equilibrium part of the free energy density according to Minerbo's model (blue) and Levermore's model (red).}
	\label{fig:freeee}
	\end{center}
\end{figure}
\begin{flalign}
\text{Minerbo:} && \label{twogangcabba7} \F &= \F_{\text{eq}}+\bigg[ \dfrac{5}{6}\bigg( \dfrac{v_{\text{ch}}w}{T}\bigg)^2-\dfrac{3}{8} \bigg( \dfrac{v_\text{ch} w}{T}\bigg)^4 +\dfrac{1}{6} \bigg( \dfrac{v_\text{ch} w}{T}\bigg)^6-\dfrac{1}{48} \bigg( \dfrac{v_\text{ch} w}{T}\bigg)^8 \bigg] \mathfrak{R} \;,&\\
\text{Levermore:} && \label{threegangcabba7} \F &= \F_{\text{eq}} +\bigg[\dfrac{5}{12} \bigg( \dfrac{v_{\text{ch}}w}{T}\bigg)^2+\dfrac{1}{36}\bigg( \dfrac{v_\text{ch} w}{T}\bigg)^6\bigg] \mathfrak{R} \;.
\end{flalign}
These are the equations of state we were looking for.

\section{Conclusions}\label{conclu}

At present, we don't know whether the GM theory is applicable outside of the Israel-Stewart regime. For that to happen, three conditions need to be met. First, the dynamics of heat must be fully characterized by a single non-equilibrium structural variable $w_\mu$. Secondly, it should still be possible to define an extended thermodynamic theory involving the non-equilibrium excursion parameter $w^\mu w_\mu +T^2$. Finally, the equation of motion for $w_\mu$ should be governed by GENERIC dynamics. And all of this must be true for large values of heat flux. Admittedly, these are quite strong assumptions to digest. However, given the success of GENERIC in describing complex fluids \cite{OttingerReview2018}, it may happen that certain relativistic liquids indeed fulfill the requirements.

The main danger when dealing with ``far-from-equilibrium'' theories of this kind is the risk of non-falsifiability. There is so much freedom in the construction of the non-equilibrium equation of state $\F(T,n,w^\mu w_\mu)$ that it is virtually possible to fit any given data a posteriori, by simply adjusting the equation of state to the needs. This would likely result in overfitted fluid models. To avoid this problem, one should know the non-equilibrium equation of state \textit{before} fitting the data with hydrodynamics. Ideally, the (theoretical) error bars of the non-equilibrium equation of state should be smaller than the (experimental) error bars of the data points.

Here, we have proposed a simple procedure for evaluating $\F(T,n,w^\mu w_\mu)$ for any given microscopic model. This procedure has the advantage of being free of intrinsic uncertainties. Rather than coming up with a statistical interpretation of $\F$, which would suffer from ambiguities related to the unclear microscopic definition of $w_\mu$, we adopted a more rigorous approach: We showed that $T$, $w^\mu w_\mu$, and $\partial \F/\partial (w^\mu w_\mu)$ can all be expressed in terms of physical observables that are \textit{unambiguously} defined in any kinetic theory. Therefore, if the GM theory holds for large values of $q$ (which is admittedly a big ``if''), there is one and only one free energy for each given microscopic model. This makes the GM theory at least falsifiable.

We have tested the method with two simple kinetic models of flux-limited diffusion. The first, due to Minerbo \cite{MINERBO1978}, based on the maximum entropy principle, and the second, due to Levermore \cite{Levermore1979,Levermore1981}, based on the Chapman-Enskog procedure \cite{ChapmanCowling}. The resulting non-equilibrium free energies are reported in figure \ref{fig:freeee}. Their qualitative behavior is reasonable. For example, the non-equilibrium deviations of the free energy are of the order of the pressure anisotropy, which is rather natural, considering that $\F=\mu n-P$. Indeed, all the scaling laws agree with the expectations.

\section*{Acknowledgements}

L.G. is partially supported by a Vanderbilt's Seeding Success Grant. This research was supported in part by the National Science Foundation under Grant No. PHY-1748958

\appendix

\section{Microscopic derivation of the toy model}\label{AAA}

\subsection{Basic definitions}

Our model \ref{ortega} describes heat conduction in a quantum liquid, which exhibits different types of weakly interacting elementary excitations (called ``branches''), that form an ideal gas. Each type of elementary excitation can be viewed as a quasi-particle \footnote{We work with quasi-particles rather than with conventional particles because we are mostly interested in applying the GM framework to neutron-star matter and quark matter. Both can be viewed as quantum liquids, and the quasi-particle picture was proven effective in both contexts \cite{chamelhaensel2006,Nakano2020,Liu:2023kzy,Bluhm:2004xn,Mykhaylova:2020pfk,Li:2022ozl}.} \cite{Arteaga:2008ux}, and it has an associated dispersion relation. The main working assumption of the model is that all the heat and the pressure anisotropy is carried by a quasi-particle ``$Q$'' that transports zero net baryon number and has a long mean free path (similarly to what happens in radiation hydrodynamics \cite{Weinberg1971,UdeyIsrael1982}). Thus, if $f_\textbf{p}$ is the distribution function of the quasiparticles, we have the following formulas (in the rest frame of $u^\mu$) \cite{landau9,landau10,Popov:2006nc}:
\begin{equation}
\begin{split}
q^j={}& \int \dfrac{d^3 p}{(2\pi)^3} f_{\textbf{p}} \, \mathfrak{q}^j \, ,\\
\Delta P^{jk} ={}& \int \dfrac{d^3 p}{(2\pi)^3} f_{\textbf{p}} \, \mathfrak{p}^{jk} \, , \\
\end{split}
\end{equation}
where $q^j$ is the heat flux, and $\Delta P^{jk}$ is the contribution to the stress tensor coming from the quasi-particles $Q$. The vector $\mathfrak{q}^j$ and tensor $\mathfrak{p}^{jk}$ are the contributions to respectively the heat flux and the stress tensor coming from a single excitation with momentum $p^j$. By isotropy of the background state, we have that $\mathfrak{q}^j \propto p^j$ and $\mathfrak{p}^{jk} \propto p^j p^k$. 

It is natural, when working within the $M1$ closure scheme, to assume that $f_{\textbf{p}}=f(\Omega)f(\epsilon_p)$, where $f(\Omega)$ is the normalized angle distribution of the quasi-particles ($\Omega$ is a point on the two-sphere), and $f(\epsilon_p)$ quantifies how many quasiparticles have energy $\epsilon_p$. Therefore, we can split the momentum integral into two separate integrals:
\begin{equation}
\begin{split}
q^j={}&  \int_{4\pi} \dfrac{d^2\Omega}{4\pi} \,  f(\Omega)\Omega^j \int_0^{+\infty} \dfrac{p^2 dp}{2\pi^2} f(\epsilon_p) \mathfrak{q} \, ,\\
\Delta P^{jk} ={}& \int_{4\pi} \dfrac{d^2\Omega}{4\pi} \, f(\Omega)\Omega^j \Omega^k   \int_0^{+\infty} \dfrac{p^2 dp}{2\pi^2} f(\epsilon_p) \mathfrak{p}  \, , \\
\end{split}
\end{equation}
where $d^2\Omega$ is solid-angle element, $\Omega^j$ is the unit vector pointing in the direction of $p^j$. Let us define the scalars
\begin{equation}\label{rererrresbuf}
    \mathfrak{R}= \int_0^{+\infty} \dfrac{p^2 dp}{2\pi^2} f(\epsilon_p) \mathfrak{p} \, , \spc v_{\text{ch}}= \dfrac{\displaystyle\int_0^{+\infty} \dfrac{p^2 dp}{2\pi^2} f(\epsilon_p) \mathfrak{q} }{\displaystyle\int_0^{+\infty} \dfrac{p^2 dp}{2\pi^2} f(\epsilon_p) \mathfrak{p}} \, .
\end{equation}
The first is just the trace $\Delta P\indices{^j _j}$. The interpretation of the second as a velocity comes from the fact that, in the non-relativistic limit, $\mathfrak{q} \sim \epsilon v \sim p v^2$, while $\mathfrak{p} \sim pv$ \cite{landau10}. However, we remark that, in relativity, $q^j$ may carry also some rest mass, and may be much larger than $\mathfrak{p}$. Thus, relativistic effects may render $v_{\text{ch}}$ larger than one.

Finally, we recall that, in the setting outlined in section \ref{genna}, the system is invariant under rotations around the $x^1$ axis. Therefore, setting up spherical coordinate such that $\Omega^1=\cos \theta$, we can write $f(\Omega)=2f(\cos \theta)$ (the 2 is a normalization constant due to the change of variables), and we obtain
\begin{equation}
    \begin{split}
q^j={}& v_{\text{ch}} \mathfrak{R} \, \delta^j_1
\int_{-1}^{+1} d(\cos \theta) \, f(\cos\theta) \cos \theta  \, , \\
\Delta P^{jk}={}& \mathfrak{R} \int_{-1}^{+1} d(\cos \theta) \, f(\cos\theta) \begin{bmatrix}
\cos^2 \theta & 0 & 0 \\
0 & \dfrac{1-\cos^2\theta}{2} & 0 \\
0 & 0 & \dfrac{1-\cos^2\theta}{2} \\
\end{bmatrix} \, . \\ 
    \end{split}
\end{equation}
Comparison with \eqref{struzzi} gives $q=q^1$, and $P_L-P_T=\Delta P^{11}-\Delta P^{22}$, so that 
\begin{equation}\label{finalmenteancora}
    \begin{split}
& q= v_{\text{ch}} \, \mathfrak{R} \int_{-1}^{+1} d(\cos \theta) f(\cos\theta) \, \cos \theta \, , \\
& P_L{-}P_T= \mathfrak{R} \int_{-1}^{+1} d (\cos\theta) \, f(\cos\theta) \, \dfrac{3(\cos \theta)^2{-}1}{2} \,  , \\
    \end{split}
\end{equation}
in agreement with equation \eqref{finalmente}.

\subsection{Non-equilibrium temperature}

If we fix the direction of the heat flux, the hydrodynamic state-space depends on three scalars, namely $\{T,n,w^\mu w_\mu\}$, see section \ref{noneqther}. Thus, also the distribution $f_{\textbf{p}}=f(\Omega)f(\epsilon_p)$ should depend on 3 kinetic parameters. In general, we expect the angular distribution $f(\Omega)$ to depend on a single parameter $Z$, which tells us how ``anisotropic'' the gas is. The energy distribution $f(\epsilon_p)$, instead, should depend on the density $n$ (since the dispersion relation is density-dependent), and on an additional parameter $W$, which quantifies how much energy is stored in the quasi-particle branch $Q$. In principle, the detailed structure of $f(\epsilon_p)$ may depend on $Z$, too. However, this should not affect the integrals in \eqref{rererrresbuf} appreciably, if $W$ is carefully defined. Thus, we can rewrite \eqref{finalmenteancora} in the form
\begin{equation}\label{nonancora}
    \begin{split}
& q= v_{\text{ch}}(W,n)\mathfrak{R}(W,n)M_1(Z) \, , \\
& P_L-P_T = \mathfrak{R}(W,n)M_2(Z) \, , \\
    \end{split}
\end{equation}
where $M_1$ and $M_2$ are the angular integrals in \eqref{finalmenteancora}. Under these assumptions, we can write a similar formula for the entropy production rate:
\begin{equation}\label{recquies}
    \tau \nabla_\mu s^\mu = \mathfrak{S}(W,n)L(Z) \, .
\end{equation}
To derive the above equation, one can work in the relaxation-time approximation, see \cite{cercignani_book}, equation (8.16), and consider that, in our model setup, most of the dissipation is due to the anisotropy of $f_{\textbf{p}}$, so that
\begin{equation}
    \tau \nabla_\mu s^\mu = \int \dfrac{d^3 p}{(2\pi)^3} f_{\textbf{p}} \, \mathfrak{s}[f_{\textbf{p}}/f_{\text{eq},p}] \approx \int \dfrac{d^3 p}{(2\pi)^3} f_{\textbf{p}} \, \mathfrak{s}[f(\Omega)] = \int_{0}^{+\infty} \dfrac{p^2 dp}{2\pi^2} f(\epsilon_p)\int_{4\pi} \dfrac{d^2 \Omega}{4\pi} \, f(\Omega)\mathfrak{s}[f(\Omega)]  \, ,
\end{equation}
where $\mathfrak{s}=\mathfrak{s}[f_{\textbf{p}}/f_{\text{eq},p}]$ is the entropy production per quasi-particle at given $f_{\textbf{p}}/f_{\text{eq},p}$ (with $f_{\text{eq},p}$ being the equilibrium distribution), and it depends on the quantum statistics of the quasi-particles. Thus, we recover \eqref{recquies}, with
\begin{equation}
\mathfrak{S}(W,n)= \int_0^{+\infty} \dfrac{p^2 dp}{2\pi^2} f(\epsilon_p) \, , \spc 
L(Z)=  \int_{4\pi} \dfrac{d^2 \Omega}{4\pi} \, f(\Omega) \mathfrak{s}[f(\Omega)]   \, .
\end{equation}
Hence, from equation \eqref{silo}, we obtain a formula for the non-equilibrium temperature:
\begin{equation}
    T= \dfrac{\mathfrak{R}(W,n)M_2(Z)}{\mathfrak{S}(W,n)L(Z)} \, .
\end{equation}
The second (and most delicate) working assumption of the model is that the quotient $M_2(Z)/L(Z)$ does not depend on $Z$. If that is true, then $T=T(W,n)$, which can be inverted, giving $W=W(T,n)$. As a consequence, $v_{\text{ch}}=v_{\text{ch}}(T,n)$ and $\mathfrak{R}=\mathfrak{R}(T,n)$, and \eqref{nonancora} becomes \eqref{finalmente}.

\subsection{Minerbo closure}

Now we only need a formula for the dependence of $f(\Omega)$ on $Z$. Minerbo \cite{MINERBO1978} adopted Jaynes' interpretation of the entropy as missing information \cite{Jaynes1}, and postulated that the most probable distribution $f(\Omega)$, for the given value of $q^j$, is the one that maximizes the (Boltzmann) entropy at fixed $q^j$. Working with Maxwell-Boltzmann statistics \cite{huang_book,landau5}, we should therefore require that
\begin{equation}
    \delta \! \! \int_{4\pi} \dfrac{d^2 \Omega}{4\pi} (f\ln f-f-Z_j  f\Omega^j-\lambda f) =0 \, ,
\end{equation}
for any linear variation $\delta f$, at fixed values of the Lagrange multipliers $Z_j$ (constraining $q^j$) and $\lambda$ (constraining the normalization). After evaluating $\lambda$ explicitly, to guarantee that $f(\Omega)$ is indeed normalised, we obtain
\begin{equation}
    f(\Omega) = \dfrac{|\textbf{Z}| e^{Z_j \Omega^j} }{\sinh(|\textbf{Z}|)} \, .
\end{equation}
Setting $Z_j=(Z,0,0)$, and recalling that $f(\Omega)=2f(\cos \theta)$, we recover equation \eqref{twogang}.

\subsection{Levermore closure}

The approach of Levermore \cite{Levermore1979,Levermore1981} is different. The main idea is to approximately solve the Boltzmann equation in the relaxation-time approximation:
\begin{equation}
   \dfrac{d}{d l} f_{\textbf{p}} +f_{\textbf{p}}=f_{\text{eq},p} \, , 
\end{equation}
where $d/dl$ denotes the derivative along the path of the quasi-particle in phase space \cite{landau10}, parameterized in units of the relaxation time. As usual, it is assumed that $f_{\textbf{p}}=f(\epsilon_p)f(\Omega)$. Additionally, one assumes that $df(\Omega)/dl \approx 0$, because $f(\Omega)$ is expected to be a slowly varying function (compared to $\tau$). This gives
\begin{equation}\label{macheroba}
    f(\Omega) = \dfrac{f_{\text{eq},p}}{f(\epsilon_p)+df(\epsilon_p)/dl} \, .
\end{equation}
Recalling that we are working in the local rest frame defined by $u^\mu$, we see that all the angular dependence of $f(\Omega)$ is encoded in the term $df(\epsilon_p)/dl$, because both $f_{\text{eq},p}$ and $f(\epsilon_p)$ are isotropic. Furthermore, it is straightforward to see that the dependence of $df(\epsilon_p)/dl$ on the vector $\Omega$ comes from terms proportional to $\Omega^j \partial_j F(\epsilon_p,T,n)$, for some function $F$. Thus, equation \eqref{macheroba} can be rearranged in the abstract form
\begin{equation}\label{sonostanco}
    f(\Omega) = \dfrac{1}{\lambda-Z_j \Omega^j} \, ,
\end{equation}
where $\lambda$ and $Z_j$ do not depend on $\Omega$. Recalling that $f(\Omega)$ is normalised, equation \eqref{sonostanco} is equivalent to
\begin{equation}
    f(\Omega)= \dfrac{1}{|\textbf{Z}|\coth(|\textbf{Z}|)-Z_j \Omega^j} \, .
\end{equation}
Setting $Z_j=(Z,0,0)$, and recalling that $f(\Omega)=2f(\cos \theta)$, we recover equation \eqref{threegang}.

\bibliography{Biblio}

\label{lastpage}

\end{document}